\documentclass[10pt,preprint]{aastex}

\slugcomment{version 2010.09.04 accepted}

\shorttitle{Star Forming Dense Cloud Cores in the TeV $\gamma $-ray SNR RX J1713.7-3946}
\shortauthors{Sano et al.}

\begin{document}


\title{Star Forming Dense Cloud Cores in the TeV $\gamma $-ray SNR RX J1713.7-3946}


\author{H. Sano\altaffilmark{1}, J. Sato\altaffilmark{1}, H. Horachi\altaffilmark{1}, N. Moribe\altaffilmark{1}, H. Yamamoto\altaffilmark{1}, T. Hayakawa\altaffilmark{1}, K. Torii\altaffilmark{1}, A. Kawamura\altaffilmark{1},\\ T. Okuda\altaffilmark{1}, N. Mizuno\altaffilmark{1,2}, T. Onishi\altaffilmark{1,3}, H. Maezawa\altaffilmark{4}, T. Inoue\altaffilmark{2}, S. Inutsuka\altaffilmark{1}, T. Tanaka\altaffilmark{5}, H. Matsumoto\altaffilmark{1},\\ A. Mizuno\altaffilmark{4}, H. Ogawa\altaffilmark{3}, J. Stutzki\altaffilmark{6}, F. Bertoldi\altaffilmark{7}, S. Anderl\altaffilmark{7}, L. Bronfman\altaffilmark{8}, B. -C. Koo\altaffilmark{9}, M. G. Burton\altaffilmark{10},\\ A. O. Benz\altaffilmark{11} and Y. Fukui\altaffilmark{1}}
\affil{$^1$Department of Physics and Astrophysics, Nagoya University, Chikusa-ku, Nagoya, Aichi, 464-8602, Japan; sano@a.phys.nagoya-u.ac.jp}
\affil{$^2$National Astronomical Observatory of Japan, Mitaka, Tokyo, 181-8588, Japan}
\affil{$^3$Department of Astrophysics, Graduate School of Science, Osaka Prefecture University, 1-1 Gakuen-cho, Nakaku, Sakai, Osaka 599-8531, Japan}
\affil{$^4$Solar-terrestrial Environment Laboratory, Nagoya University, Furo-cho, Chikusa-ku, Nagoya, Aichi 464-8601, Japan}
\affil{$^5$Kavli Institute for Particle Astrophysics and Cosmology, Stanford University, 382 Via Pueblo Mall, MC 4060, Stanford, CA 94305}
\affil{$^6$KOSMA, I. Physikalisches Institut, Universit$\ddot{a}$t zu K$\ddot{o}$ln, Z$\ddot{u}$lpicher Str$\ddot{a}$sse 77, 50937 K$\ddot{o}$ln, Germany}
\affil{$^7$Argelander-Institut f$\ddot{u}$r Astronomie, Auf dem H$\ddot{u}$gel 71, 53121 Bonn, Germany}
\affil{$^{8}$Departamento de Astronomia, Universidad de Chile, Casilla 36-D, Santiago, Chile}
\affil{$^{9}$Seoul National University, Seoul 151-742, Korea}
\affil{$^{10}$School of Physics, University of New South Wales, Sydney, NSW 2052, Australia}
\affil{$^{11}$Institute of Astronomy, ETH Z$\ddot{u}$rich, 8093 Z$\ddot{u}$rich, Switzerland}





\begin{abstract}
RX J1713.7-3946 is one of the TeV $\gamma$-ray supernova remnants (SNRs) emitting synchrotron X rays. The SNR is associated with molecular gas located at $\sim $1 kpc. We made new molecular observations toward the dense cloud cores, peaks A, C and D, in the SNR in the $^{12}$CO($J$=2-1) and $^{13}$CO($J$=2-1) transitions at angular resolution of 90". The most intense core in $^{13}$CO, peak C, was also mapped in the $^{12}$CO($J$=4-3) transition at angular resolution of 38". Peak C shows strong signs of active star formation including bipolar outflow and a far-infrared protostellar source and has a steep gradient with a $r^{-2.2\pm 0.4}$ variation in the average density within radius $r$. Peak C and the other dense cloud cores are rim-brightened in synchrotron X rays, suggesting that the dense cloud cores are embedded within or on the outer boundary of the SNR shell. This confirms the earlier suggestion that the X rays are physically associated with the molecular gas (Fukui et al. 2003). We present a scenario where the densest molecular core, peak C, survived against the blast wave and is now embedded within the SNR. Numerical simulations of the shock-cloud interaction indicate that a dense clump can indeed survive shock erosion, since shock propagation speed is stalled in the dense clump. 
Additionally, the shock-cloud interaction induces turbulence and magnetic field amplification around the dense clump that may facilitate particle
acceleration in the lower-density inter-clump space leading to the enhanced synchrotron X rays around dense cores.


\end{abstract}

\keywords{ISM: molecules --- ISM: individual objects (RX J1713.7-3946) --- ISM: supernova remnants --- Stars: protostars --- X-rays: ISM}

\section{Introduction}

Supernova remnants (SNRs) have a profound influence on the interstellar medium (ISM) via shock interaction and injection of heavy elements. The dynamical interaction also affect the evolution of SNRs through the distortion of the shell morphology, if the ISM is dense enough. It is therefore important to study the detailed physical properties in the interaction between SNRs and ISM, in oder to understand what can occur.

RX J1713.7-3946 (G347.3-0.5) is one of the high energy $\gamma $ ray SNRs associated with dense molecular gas and is a good target to study the interaction. The $ROSAT$ satellite detected the SNR at X rays for the first time (Pfeffermann $\&$ Aschenbach 1996). Koyama et al. (1997) showed that the X ray emission is synchrotron emission without thermal features using the $ASCA$ satellite and derived a distance of 1kpc based on a relatively small X ray absorption, corresponding to an HI column density of (6.2$\pm $1)$\times $10$^{21}$ cm$^{-2}$ for the northwestern rim. Slane et al. (1999), on the other hand, claimed a distance of 6 kpc based on three CO clouds of around $V_\mathrm{LSR}$ of -95 km s$^{-1}$ observed at 8.8 arcmin resolution, which were suggested to be interacting with the SNR by the authors. They argued that the small X ray absorption is ascribed to an HI hole that happens to be located toward the SNR.

Subsequently, Fukui et al. (2003) revealed that molecular gas at a much lower $V_\mathrm{LSR}$ of -11 km s$^{-1}$ to -3 km s$^{-1}$ is interacting with the SNR and derived a kinematic distance of 1 kpc by using the NANTEN CO dataset at a higher angular resolution of 2.6 arcmin (Mizuno $\&$ Fukui 2004). Figure 1 shows the CO distribution in this velocity range overlayed with the X ray distribution, and demonstrates that the molecular gas shows a clear anti-correlation with the X ray distribution and that intense molecular peaks are spatially correlated with the X ray peaks. Observations with the $XMM-Newton$ satellite derived a distance around 1.3 kpc from a study of X ray absorption at a higher angular resolution than $ASCA$ (Cassam-Chenai et al. 2004). Moriguchi et al. (2005) showed further details of the $^{12}$CO($J$=1-0, 3-2) distributions and confirmed the association between the SNR and the low velocity molecular gas. Aharonian et al. (2004, 2006 and 2007) revealed a shell like TeV $\gamma $ distribution with the H.E.S.S. Cerenkov $\gamma $ ray telescope and showed that the SNR is a very energetic emitter of TeV $\gamma $ rays. They compared the $\gamma $ rays with the NANTEN CO distribution and showed that the CO and $\gamma $ rays exhibit a fairly good correlation particularly toward the molecular peaks in the northwestern rim, supporting the physical association between the SNR and the molecular gas. Tanaka et al. (2008) presented a detailed comparison of $Suzaku$ X ray data with TeV $\gamma $ rays and discussed the origin of the $\gamma $ rays, whereas the connection between high energy photons and the molecular gas remained unexplored. To summarize, RX J1713.7-3946 shows a close connection among X rays, $\gamma $ rays and molecular gas, a unique case where high energy particles, either protons or electrons, and the low energy ISM are co-existent and physically interacting (Fukui et al. 2008). 

In the course of the above studies Fukui et al. (2003) found that CO peak C shows broad CO wings and suggested that the wings may result from dynamical acceleration by the SNR blast wave. Such broad molecular wings are found in several SNRs including IC443, W44, W28, etc. (e.g., Denoyer et al. 1979, Wootten 1977, 1981). Moriguchi et al. (2005) showed that the $^{12}$CO($J$=3-2) distribution in peak C shows a hint of a bipolar nature being associated with an infrared compact source with the spectrum of a protostar. This may indicate an alternative possibility that the broad wings are driven by the outflow from a protostar but are not due to the shock interaction. It is an open question whether the SNR is accelerating molecular gas to high velocities in peak C. It is therefore important to clarify whether the CO broad wings are due to the blast-wave acceleration or due to protostellar activity in our efforts to better understand the interaction.   

We have carried out new observations of the molecular peaks associated with the X rays in the $^{12}$CO($J$=2-1, 4-3) and $^{13}$CO($J$=2-1) transitions at mm and sub-mm wavelengths in order to derive the molecular properties of the core and broad wings in peak C and the other cloud cores nearby in RX J1713.7-3946. We present these molecular results and discuss star formation in the SNR in this paper. We will publish separate papers that deal with a detailed X ray analysis of $Suzaku$ results, a comparison with TeV $\gamma $ rays and an analysis of the $Spitzer$ results toward the molecular peaks. In the following we shall adopt a distance 1 kpc and assume that the SNR has an age of 1600 yrs, a radius of 7 pc and an expansion speed of 3000 km s$^{-1}$ (Uchiyama et al. 2007). This is consistent with the SNR corresponding to a historical SNR recorded in the Chinese literature in AD393/4 (Wang et al. 1997). 

The present paper is organized as follows. Section 2 describes the observations and Section 3 their results. Section 4 gives an analysis of density and temperature. The paper is summarized in Section 5.

\section{Observations}

We carried out $^{12}$CO($J$=2-1, 4-3) and $^{13}$CO($J$=2-1) observations with the NANTEN2 4 m sub-mm telescope of Nagoya University installed at Pampa La Bola (4865 m above the sea level) in the northern Chile.

Observations in the $^{12}$CO($J$=2-1) and $^{13}$CO($J$=2-1) lines were conducted from August to November and in December 2008, respectively. The backend was a 4 K cooled Nb SIS mixer receiver and the single-side-band (SSB) system temperature was $\sim $250 K for both transitions, including the atmosphere toward the zenith.  The telescope had a beam size of 90" at 230 GHz. The pointing was checked by observing the Jupiter every two hours and was found to be as accurate as $\sim $15". We used acoustic optical spectrometers (AOS) with 2048 channels having a bandwidth of 390 km s$^{-1}$ and resolution of 0.38 km s$^{-1}$. Observations in $^{12}$CO were carried out in the on-the-fly (OTF) mode, scanning with an integration time of 1.0 to 2.0 sec. The chopper wheel method was employed for the intensity calibration and the derived Trms was better than 0.66 and 0.51 K ch$^{-1}$ with 1.0 and 2.0 sec integrations, respectively. The observed area was about 2.25 square degrees and is shown in Figure 1. Observations in $^{13}$CO were also carried out in the on-the-fly (OTF) mode for the area of 22 square arcminutes including the peaks A, B, C, and D (Moriguchi et al. 2005) with an integration time of 2.0 sec with as Trms better than 0.68 K ch$^{-1}$. The absolute intensity was calibrated by observing Oph EW4 [$16^{\mathrm{h}} 26^{\mathrm{m}} 21\fs92; -24{^\circ}  25\arcmin 40\farcs4$ (J2000)] (Kulesa et al. 2005). 

Observations in $^{12}$CO($J$=4-3) were carried out from November to December 2007 covering a 9 square arcminutes region including peak C and toward a point in peak A. The telescope had a beam size of $38"$ at 460 GHz as measured by observing Jupiter. The front end was a SIS receiver having SSB temperature of 300 K including the atmosphere toward the zenith. The typical rms noise fluctuations were 0.28 K ch$^{-1}$. The absolute intensity calibration is as described by Pineda et al. (2008).

Together with the data above, we made use of the data by Moriguchi et al. (2005). They observed the peaks A, C, and D with the ASTE sub-mm telescope in $^{12}$CO($J$=3-2) in November, 2004. The data were taken by a position switching mode with a 30" grid spacing with 23" beam. The spectrometer was an AOS with 450 km s$^{-1}$ band width and 0.43 km s$^{-1}$ resolution. The system temperature was 300 - 400 K (DSB) and the typical rms achieved is 0.4 - 0.9 K with 30 sec integration.

\section{Results}

\subsection{$^{12}$CO($J$=2-1) and $^{13}$CO($J$=2-1) distributions}

Figures 2a and 2b show the $^{12}$CO($J$=2-1) and $^{13}$CO($J$=2-1) distributions, respectively, of the western rim of the SNR in the box shown by dashed lines in Figure 1. There are nearly ten CO peaks and the observational parameters of nine of them are given in Table 1 as taken from Fukui et al. (2003) and Moriguchi et al. (2005). Three of the cores marked by crosses in Figure 2a, peaks A, C and D, are associated with $IRAS$ point sources with a protostellar spectral energy distribution (SED) (Moriguchi et al. 2005). Bright cores in $^{12}$CO are all associated with $IRAS$ sources and we mapped them in the $^{13}$CO($J$=2-1) transition for the two areas shown by solid lines in Figure 2. All the cores with $IRAS$ sources are detected in the $^{13}$CO transition and peak C is the most intense among them.

\subsection{$^{12}$CO($J$=4-3) distribution}

We observed the $^{12}$CO($J$=4-3) transition in an area of 3 arcmin $\times $ 3 arcmin around peak C in equatorial coordinates and toward the peak position of peak A. Figure 3 shows four images of peak C in the $^{12}$CO($J$=2-1, 3-2, and 4-3) and $^{13}$CO($J$=2-1) transitions, where the $^{12}$CO($J$=3-2) distribution is taken from Moriguchi et al. (2005). The $^{12}$CO($J$=4-3) core is most compact and the size of the core increases toward the lower $J$ transitions, suggesting a sharp density decrease with radius, since the higher $J$ transitions have higher critical densities for collisional excitation.

\subsection{$^{12}$CO($J$=4-3) broad wings}

Figure 4 shows that the broad $^{12}$CO($J$=4-3) wings first detected by Fukui et al. (2003) reveal a clear bipolar signature centered on $IRAS$ 17089-3951 (Table 2) and on the peak position of the dense cloud core in peak C. The bipolarity verifies that the wings are driven by a protostar and are not driven by the SNR blast wave. The $IRAS$ source position also shows a good correlation with an extended MSX sources at 8.28 $\mu $m (from IPAC Infrared Science Archive) (Figure 4a).
The wings toward peak C are also recognized in the present $^{12}$CO($J$=2-1) data in addition to the $^{12}$CO($J$=1-0) and $^{12}$CO($J$=3-2) transitions (Moriguchi et al. 2005), whereas the wing intensities of these lower-$J$ transitions are more than a few times weaker than the $^{12}$CO($J$=4-3) wings. It is not clear if the other peaks A and D show sings of bipolar outflow either in the $^{12}$CO($J$=1-0) and $^{12}$CO($J$=3-2) data (Moriguchi et al. 2005) or in the present $^{12}$CO($J$=2-1) data.

\subsection{Comparison with X rays}

Figure 5(a) shows an overlay of the $^{12}$CO($J$=2-1) image with the $Suzaku$ X ray image in the 1-5keV range over the western rim of the SNR (Tanaka et al. 2008). As noted earlier in Moriguchi et al. (2005), in the lower resolution $^{12}$CO($J$=1-0) image at 2.6 arcmin resolution, and the $^{12}$CO($J$=3-2) distributions of more limited coverage at 30 arcsec resolution for individual peaks A, C and D, the X ray distribution shows a clear correlation with the molecular distribution. First, peak C is surrounded by bright X ray emission both on its north and south with a local minimum toward the center of the core. Figure 5(b) shows a remarkable coincidence between the X ray depression and the cloud core at a $\sim $0.1 pc scale. This depression is not due to interstellar absorption because we find similar distribution of X ray in the higher energy band (5-10 keV) that is hardly absorbed (Figure 5(c) and (d)). Therefore, this morphology suggests that the X ray emission is enhanced on the surface of the cloud core. Peak D is also associated with bright X ray emission, whereas only the eastern part, the inner side of the shell, is emitting the X rays. This is ascribed to that peak D is located on the outer surface of the SN shell and only the inner side is being illuminated by the shock. Another X ray feature is toward peaks A and B where the X rays are bright between the two peaks, showing a similar anti-correlation. Such a trend is also seen in the middle of the SNR rim between peaks E and G and between peaks E and I. All these features suggest that the X ray emission is enhanced on the surface of the molecular gas.

\section{Analysis of molecular properties}

\subsection{LVG analysis}

We shall use the large velocity gradient model of line radiation transfer to estimate density and temperature from the muti-$J$ transitions of CO; i.e., $^{12}$CO($J$=3-2, 4-3) and $^{13}$CO($J$=2-1). We applied the large velocity gradient (LVG) analysis (Goldreich $\&$ Kwan 1974; Scoville $\&$ Solomon 1974) to estimate the physical parameters of the molecular gas toward peaks A and C by adopting a spherically symmetric uniform model having a radial velocity gradient dv/dr.
The $^{12}$CO($J$=2-1) transition was not included in the analysis because the transition may be subject to self absorption due to low excitation foreground gas (e.g. Mizuno et al. 2010). We calculate level populations of $^{12}$CO and $^{13}$CO molecular rotational states and line intensities. The LVG model requires three independent parameters to calculate emission line intensities; i.e., kinetic temperature, density of molecular hydrogen and X/(dv/dr). X/(dv/dr) is the abundance ratio from CO to H$_2$ divided by the velocity gradient in the cloud. We use the abundance ratio [$^{12}$CO]/[$^{13}$CO] $\sim $75 (G$\mathrm{\ddot{u}}$sten et al. 2004) and [$^{12}$CO]/[H$_2$] $\sim $5$\times $10$^{-5}$ (Blake et al. 1987), and estimate the mean velocity gradient between the peaks as $\sim $12.5 km s$^{-1}$ pc$^{-1}$. Accordingly, we adopt that X/(dv/dr) is 4.0$\times $10$^{-6}$ (km s$^{-1}$ pc$^{-1}$)$^{-1}$ for $^{12}$CO. 

In order to solve temperatures and densities which reproduce the observed line intensity ratio, we calculate chi-square $\chi^2$ defined as below;

\begin{eqnarray}
\chi ^2 &= & \Sigma [(R_\mathrm{obs} - R_\mathrm{model})^2/\sigma]
\end{eqnarray}

where $R_\mathrm{obs}$ is the observed line intensity ratio between different excitation lines or different isotopes, $R_\mathrm{model}$ is the line ratio of the LVG calculations, $\sigma $ is the standard deviation for $R_\mathrm{obs}$ 
in the analysis. The error in the observed intensity is estimated by considering the noise level of the observations and the calibration error. We assume that the error of calibration from $T_A^*$ to $T_{MB}$ is 10 $\%$ for all the line intensities. The data used are derived from the line profiles in Figure 6 upper and three ratios are estimated for a 1.5 km s$^{-1}$ velocity interval in the three peaks as listed in Table 3. 

The lower panels of Figure 6 shows the results of fitting to the data obtained with a chi-square minimization approach to find the solution of temperature and density. Each locus of a black solid line surrounding the cross indicates the chi-square $\chi^2$, which corresponds to the 
95$\%$ confidence level of a chi-square distribution. The crosses denote the lowest point of chi-square. 
Additionally, we are able to reject each region outside the black solid line at the 95$\%$ confidence level. 
Table 3 summarizes the results of the LVG analysis. Density and temperature are relatively well constrained in peaks A and C. 
The temperature of peaks A and C is in a range of 10 - 12 K and density is somewhat higher in peak C, $\sim $10$^4$ cm$^{-3}$, than in peak A, $\sim $6$\times $10$^3$ cm$^{-3}$.

\subsection{The density distribution of peak C}

Peak C is associated with a dense cloud core with a strong intensity gradient (Figure 3). The total molecular mass of the core is estimated to be 400 M$_{\odot}$  from the $^{12}$CO($J$=1-0) integrated intensity (Moriguchi et al. 2005) for an X factor of 2.0 $\times $ 10$^{20}$ [$W$($^{12}$CO)/(K km s$^{-1}$)](Bertsch et al. 1993). We shall derive the density distribution by employing a simple power-law analysis assuming a spherical symmetry. First, we de-convolve the intensity distribution in order to correct for the beam size. We assume the relation to derive a de-convolved core radius $r$ in each transition; $r^2$ = $s^2$ + $r'^2$, where $r$, $s$ and $r'$ stand for the observed radius, the beam radius and the de-convolved radius, respectively, at a half power level of the peak intensity (Table 4). Then, we estimate the averaged density within $r$, $\overline{n(\mathrm{H_2}) (<r)}$, so as to match the observed averaged integrated intensity to the LVG estimate within the radius by assuming kinetic temperature of 12 K and the same model parameters in Section 4.1. Considering the low luminosity of the $IRAS$ source (300 $L_{\odot}$), we infer that the local temperature variation in the core is not significant and that a uniform temperature is a good approximation. The result is shown in Figure 7. This presents that the average density distribution is well approximated by a power law, $r^{-2.2\pm 0.4}$. Such a steep density gradient is consistent with a star forming cloud core (Larson 1969, Penston 1969, Lizano and Shu 1989, Onishi et al. 1999). The line width does not vary much among the different density regimes as shown in Figure 7, suggesting the infall motion is not very large. An alternative interpretation for the density distribution will be discussed later in Section 5.

\subsection{Physical parameters of the outflow}

The bipolarity of the broad CO wings in Figure 4 verifies that the wings are caused by bipolar outflow driven by a protostar and is consistent with the compact dense core with a steep density gradient (Section 4.2). The physical parameters of the outflow are estimated by using the method described by Moriguchi et al. (2005) (Table 5). The average intensities of the red component are 8.9 K and 2.0 K for the $^{12}$CO($J$=2-1) and $^{12}$CO($J$=4-3) transitions in a velocity range from -8 km s$^{-1}$ to -3 km s$^{-1}$. The density in the line emitting wings are then estimated to be $\sim 5 \times  10^3$ cm$^{-3}$ from the line intensity ratio, $\sim $0.23, between the $^{12}$CO($J$=2-1) and $^{12}$CO($J$=4-3) transitions. The beam filling factor is estimated to be as small as 5$\%$ for an assumed excitation temperature of 12 K for the rotational levels of $^{12}$CO. This result is similarly obtained from the blue component, that yields beam filling factor of $\sim 4\%$. Therefore, the CO wings represent highly clumped gas 
rather typical for outflows. The present $^{12}$CO($J$=4-3) data have verified that the wings represent outflow and showed that the $^{12}$CO($J$=4-3) wings are much more compact than the $^{12}$CO($J$=1-0) wings, indicating that the $^{12}$CO($J$=4-3) wings are denser outflow gas whose spatial extent is smaller than the lower density wings (Table 5, for the $^{12}$CO($J$=1-0) wings see Table 4 of Moriguchi et al. 2005). The most likely driving source of the outflow is the far infrared source $IRAS$ 17089-3951. General properties for bipolar outflows are found elsewhere (e.g., Lada 1986, Fukui 1989, Fukui et al. 1993) and the present outflow is seen to be typical of an outflow associated with a low mass protostar in terms of its size and velocity. 

\subsection{An evolutionary scheme}

We present a general scenario following discussions given by Fukui et al. (2003) and Moriguchi et al. (2005). The progenitor of the SN was a high mass star formed some Myrs ago in the region, and which created a cavity in the ISM by its stellar wind. This is a star-forming region with a loose spatial association extended over a few 10 pc. The molecular peaks, including A, B, C, D, and the others are in part of the cavity wall. The total mass of the molecular complex is at least a few 1000 $M_{\odot }$ (Moriguchi et al. 2005). The star formation in the peaks A, C, and D may have been triggered by the progenitor of the SNR (Koo et al. 2008, Desai et al. 2010). The progenitor star caused the supernova explosion 1600 yrs ago, as recorded in an ancient Chinese document (Wang et al. 1996) and the blast wave expanded into the cavity. The SNR is now interacting with the molecular gas and this interacting layer is traced by the X rays, as supported by the good spatial anti-correlation with the molecular gas at a 0.1 pc scale.

The SNR with an age of 1600 yrs is still in the free expansion phase for ambient density less than 1 cm$^{-3}$ (e.g., Truelove and McKee 1999), while the expansion may experience a sudden deceleration when it hits the dense ISM wall. The interactions of such a young SNR with the interstellar medium are numerically simulated by Inoue, Yamazaki $\&$ Inutsuka (2009). These authors postulated an inhomogeneous initial density distribution in a range of 1 - 1000 cm$^{-3}$ with a speed of the blast wave of 1000 km s$^{-1}$ and an age of 1000 yrs. The density distribution is basically consistent with the observed clumped nature of the CO gas. Based on the numerical results, we are able to estimate the disturbing depth of the blast wave into the interacting medium as a function of the initial density. 
It is found that the penetrating velocity of the shock front into a clump is inversely proportional to $\sqrt{n}$, where $n$ is the density of the clump. From the results of the numerical simulations, we find that the velocity is given as $\sim $3000 km s$^{-1}$/$\sqrt{n/n_0}$, where the ambient density $n_{0}$=1 cm$^{-3}$ in the present SNR. The penetrating depth in a typical timescale of the interaction $\sim $1000 yrs is then estimated to be $\sim $0.1 pc for $n$=10$^3$ cm$^{-3}$ and $\sim $3 pc for $n$=1 cm$^{-3}$, respectively. Peak C has a present diameter of at least 0.6 pc at density of 10$^3$ cm$^{-3}$ (Figure 7) and should have not been affected significantly by the shock penetration according to the argument above, whereas the ambient lower density gas is significantly disturbed and accelerated in a scale length of the SNR radius. We, therefore, infer that peak C is able to survive against the shock wave as originally suggested by Fukui et al. (2003), while the lower density ambient gas is significantly disturbed and accelerated .

It seems essential in the interaction that the initial density distribution in and around the stellar wind cavity was highly inhomogeneous. This inhomogeneity is indeed verified by the clumpy molecular distribution in the present region. According to the theoretical results (Inoue et al. 2009), dense clumps tend to retard shock fronts. The global morphology of blast wave is ,however, not deformed significantly because the shock wave propagates through lower density inter-clump regions between dense clumps. This explains the nearly circular shape of the SNR with little observable deformation, even if the dense clumps are located inside the shell. Another implication of the numerical simulations is that the magnetic field may be amplified considerably to mG by turbulence in the interaction where the initial density is higher. Such amplified fields are able to offer an explanation of the enhanced synchrotron X ray emission, because the synchrotron emission is proportional to $B^2$ and the number density of high energy electrons. The rim-brightened X rays toward the molecular peaks are explained by such field amplification that leads also to efficient acceleration of high energy electrons in the inter-clump space.

\section{Summary}

We present millimeter and sub-millimeter spectroscopic observations of the molecular cloud cores in the TeV $\gamma $ ray SNR RX J1713.7-3946. Our main conclusions are summarized as follows;

\begin{enumerate}
\item Three of the dense cores, peaks A, C and D, in RX J1713.7-3946 are associated with $IRAS$ point sources with protostellar spectra as noted by Moriguchi et al. (2005). These cores show $^{13}$CO($J$=2-1) emission and have densities around 10$^4$ cm$^{-3}$. Peak C is the most outstanding among the three and shows strong $^{12}$CO($J$=4-3) emission.
\item The spatial distributions of the four transitions, $^{12}$CO($J$=2-1, 3-2, 4-3) and $^{13}$CO($J$=2-1), indicate that the core of peak C has a strong density gradient consistent with an average density distribution of $r^{-2.2\pm 0.4}$, where $r$ is the radius from the center of the core. The density and temperature, averaged over 90" (=0.2 pc) are 0.8 - 1.7 $\times$ 10$^4$ cm$^{-3}$ and 11 - 16 K, as derived from an LVG analysis. Peak C is also associated with a bipolar outflow as evidenced by the $^{12}$CO($J$=4-3) broad wings of at least 30 km s$^{-1}$ velocity extent. Along with the far infrared spectrum, these identify the region as a site of recent low-mass star formation within Myr. This verifies the broad wings are not produced by the shock acceleration driven by the SNR as suggested before by Fukui et al. (2003).
\item Peak A has density and temperature of 5 - 8 $\times $ 10$^3$ cm$^{-3}$ and 9 - 11 K, somewhat lower than peak C. 
The $IRAS$ sources in peaks A and D are also likely protostars, because of the high density of these peaks.
\item A morphological comparison with X rays indicates that the dense cloud cores show a good anti-correlation with the X rays, suggesting that the X rays are enhanced on the surface or surroundings of the dense molecular gas where magnetic field strength is enhanced. We show that these features are consistent with theoretical results of shock propagation in the highly inhomogeneous ISM. Numerical simulations support that peak C is a dense clump which survived against shock erosion, since shock propagation speed is stalled in the dense clump.
\end{enumerate}

\acknowledgments
The NANTEN project is based on a mutual agreement between Nagoya University and the Carnegie Institution of Washington (CIW). We greatly appreciate the hospitality of all the staff members of the Las Campanas Observatory of CIW. We are thankful to many Japanese public donors and companies who contributed to the realization of the project. NANTEN2 is an international collaboration of ten universities, Nagoya University, Osaka Prefecture University, University of Cologne, University of Bonn, Seoul National University, University of Chile, University of New South Wales, Macquarie University, University of Sydney and ETH Zurich. The work is financially supported by a grant-in-aid for Scientific Research (KAKENHI, no. 15071203, no. 21253003, and no. 20244014) from MEXT (the Ministry of Education, Culture, Sports, Science and Technology of Japan) and JSPS (Japan Society for the Promotion of Science) as well as JSPS core-to-core program (no. 17004). We also acknowledge the support of the Mitsubishi Foundation and the Sumitomo Foundation. This research was supported by the grant-in-aid for Nagoya University Global COE Program, $"$Quest for Fundamental Principles in the Universe: from Particles to the Solar System and the Cosmos$"$, from MEXT. Also, the work makes use of archive data acquired with $MSX$ and $IRAS$ data gained with Infrared Processing and Analysis Center (IPAC). The satellite internet connection for NANTEN2 was provided by the Australian Research Council. We are grateful to the anonymous referee for useful comments which helped the authors to improve the paper. LB acknowledges support from Center of Excellence in Astrophysics and Associated Technologies (PFB 06) and by FONDAP Center for Astrophysics 15010003.




\clearpage



\newpage
\begin{figure}[hbpt]
\begin{center}
\includegraphics[width=159mm,clip]{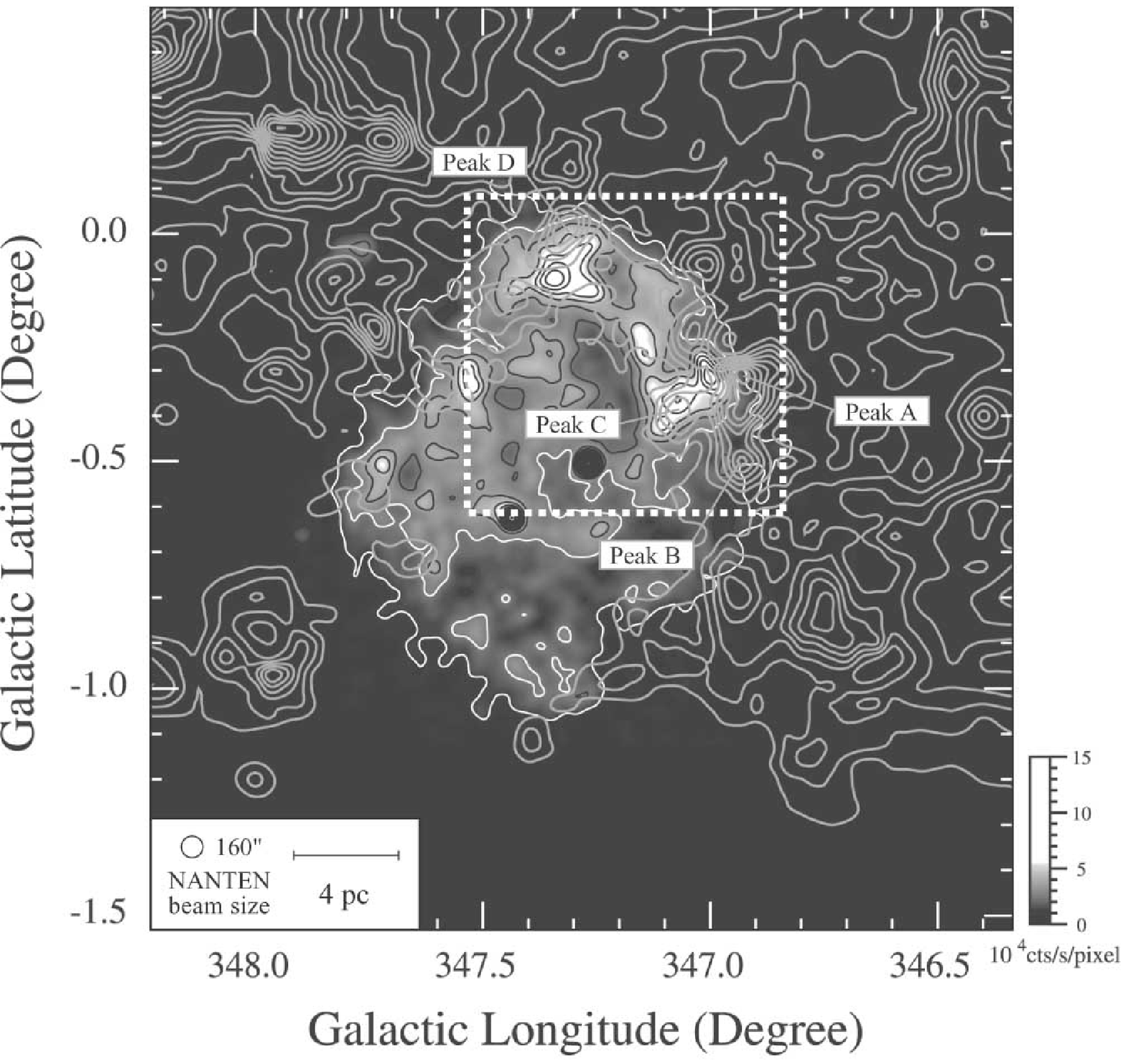}
\caption{The RX J1713.7-3946 region in galactic coordinates taken from Figure 1 of Fukui et al. (2003). The X ray distribution is shown in gray scale [$ROSAT$ PSPC X-ray Survey (Slane et al. 1999) from $ROSAT$ archive database] and the intensity distribution of $^{12}$CO($J$=1-0) emission in purple contours. The intensity is derived by integrating the CO spectra from -11 to -3 km s$^{-1}$, which corresponds to the component interacting with the SNR. The lowest contour level and contour interval are 4 K km s$^{-1}$ for each. The area enclosed by the dashed line is the observed area in the $^{12}$CO($J$=2-1) transition (see Fig.2.(a)). The major four Peaks of CO, A - D, have been used in Fukui et al. (2003).}
\label{fig1}
\end{center}
\end{figure}%

\newpage
\begin{figure}[hbpt]
\begin{center}
\includegraphics[width=162mm,clip]{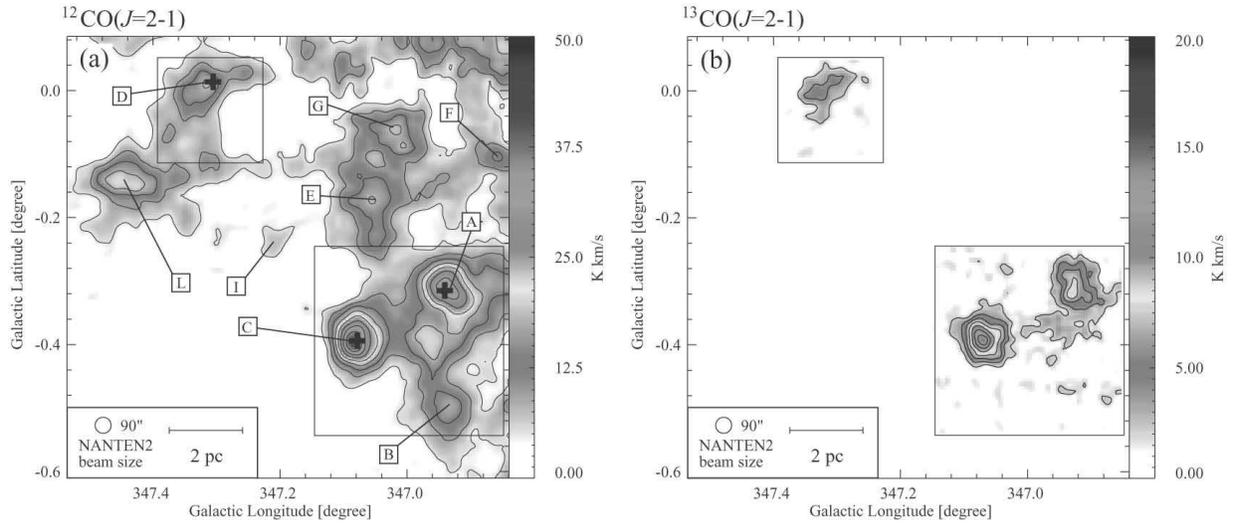}
\caption{(a) Intensity distribution of the $^{12}$CO($J$=2-1) transition. The intensity is derived by integrating the $^{12}$CO($J$=2-1) spectra from -16 to -3 km s$^{-1}$. The lowest contour level is 5.0 K km s$^{-1}$($\sim 4.2\sigma $) and the contour interval is 4.3 K km s$^{-1}$($\sim 3.6\sigma $). Open crosses show the positions of the $IRAS$ point sources (see Table 2). The area enclosed by the black line is the area observed in the $^{13}$CO($J$=2-1) transition (see Fig.2.(b)). (b) Intensity distribution of the $^{13}$CO($J$=2-1) transition. The intensity is derived by integrating the $^{13}$CO($J$=2-1) spectra from -16 to -8 km s$^{-1}$. The lowest contour level is 2.1 K km s$^{-1}$($\sim 3.1\sigma $) and the contour interval is 1.8 K km s$^{-1}$($\sim 2.6\sigma $).}
\label{fig2}
\end{center}
\end{figure}%

\newpage
\begin{figure}[hbpt]
\begin{center}
\includegraphics[width=111mm,clip]{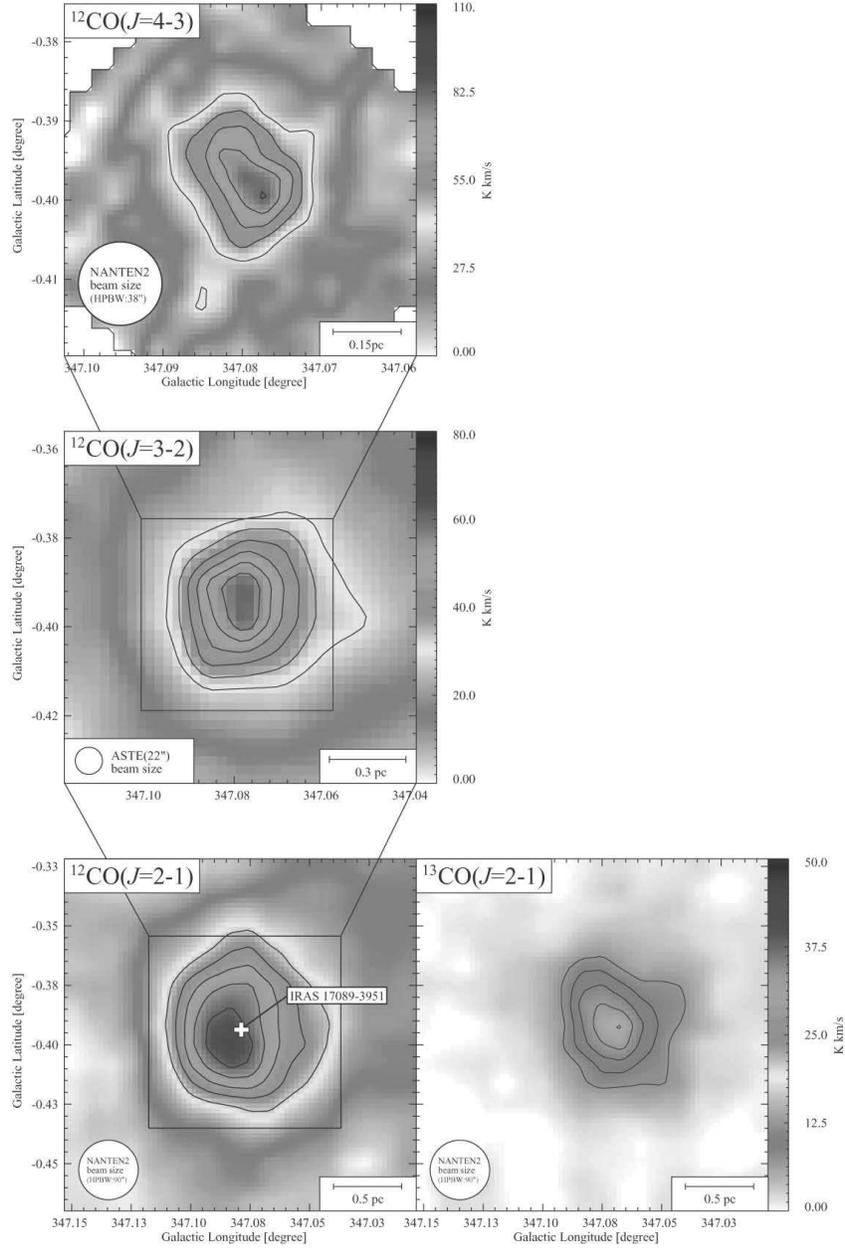}
\caption{Intensity distributions of the $^{12}$CO($J$=4-3, 3-2, 2-1) and $^{13}$CO($J$=2-1) transitions of peak C. The velocity range is from -30 to 7 km s$^{-1}$ in $^{12}$CO($J$=4-3); from -20 to -7 km s$^{-1}$ in $^{12}$CO($J$=3-2); from -16 to -3 km s$^{-1}$ in $^{12}$CO($J$=2-1); and from -16 to -8 km s$^{-1}$ in $^{13}$CO($J$=2-1). The CO contours are every 10.2 K km s$^{-1}$($\sim 3.0\sigma $) from 41.6 K km s$^{-1}$($\sim 12.2\sigma $) in $^{12}$CO($J$=4-3); every 5.0 K km s$^{-1}$ ($\sim 5.3\sigma $) from 30.2 K km s$^{-1}$ ($\sim 31.8\sigma $) in $^{12}$CO($J$=3-2); every 4.2 K km s$^{-1}$ ($\sim 3.5\sigma $) from 21.0 K km s$^{-1}$ ($\sim 17.0\sigma $) in $^{12}$CO($J$=2-1); and every 1.8 K km s$^{-1}$($\sim 2.6\sigma $) from 7.28 K km s$^{-1}$ ($\sim 10.6\sigma $) in $^{13}$CO($J$=2-1). The lowest contours are a half value of the peak intensity for each emission. The open cross shows the positions of the $IRAS$ point source (see Table 2.)}
\label{fig3}
\end{center}
\end{figure}%

\newpage
\begin{figure}[hbpt]
\begin{center}
\includegraphics[width=164mm,clip]{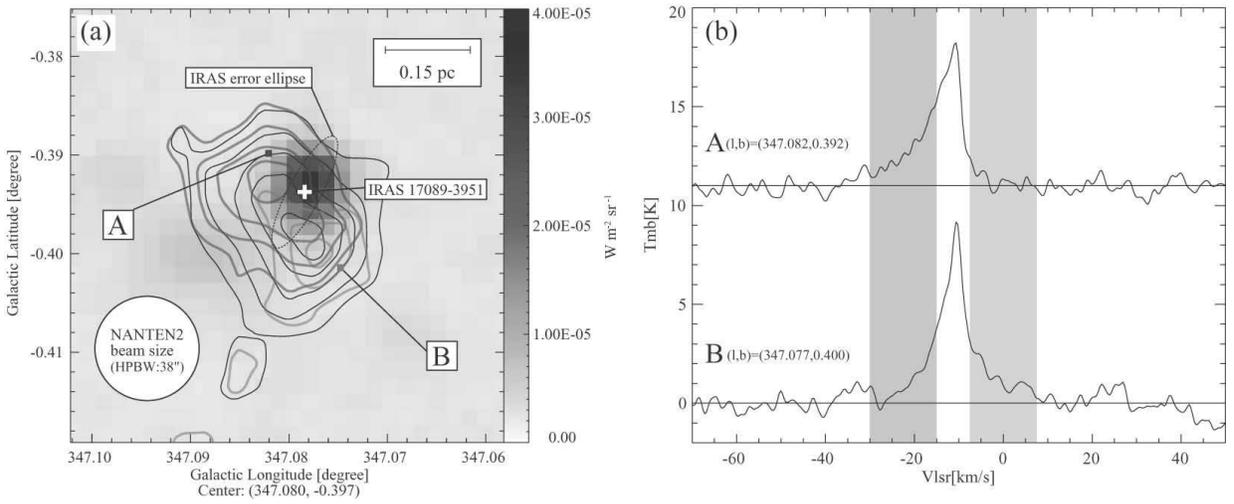}
\caption{(a) Overlay map RX J1713.7-3946 $MSX$ 8.28$\mu $m image in color scale (from IPAC Infrared Science Archive) and $^{12}$CO($J$=4-3) intensity contours taken by NANTEN2. The velocity range is from -30 to 7 km s$^{-1}$ for the black contours. The blue contours are from -30 to -15 km s$^{-1}$ and the red contours from -8 to 7 km s$^{-1}$. The lowest contour level of the black is 37.3 K km s$^{-1}$ ($\sim 11\sigma $) and the red and blue are 9.6 K km s$^{-1}$ ($\sim 4\sigma $) for each. The contour interval of the black contour is 10.2 K km s$^{-1}$ ($\sim 3\sigma $) and the others are 4.8 K km s$^{-1}$ ($\sim 2\sigma $). Open cross and enclosed black dashed circle show the positions of the $IRAS$ point source (see Table 2.) and the 90$\%$ confidence region. (b)Top: $^{12}$CO($J$=4-3) spectrum at (l,b)=(347.082deg, -0.392deg) covering a velocity range from -70 to 50 km s$^{-1}$. Bottom: $^{12}$CO($J$=4-3) spectrum at (l, b)=(347.077deg, -0.400deg) covering the same velocity range. Painted areas with blue and red correspond to those color contours in (a).}
\label{fig4}
\end{center}
\end{figure}%

\newpage
\begin{figure}[hbpt]
\begin{center}
\includegraphics[width=164mm,clip]{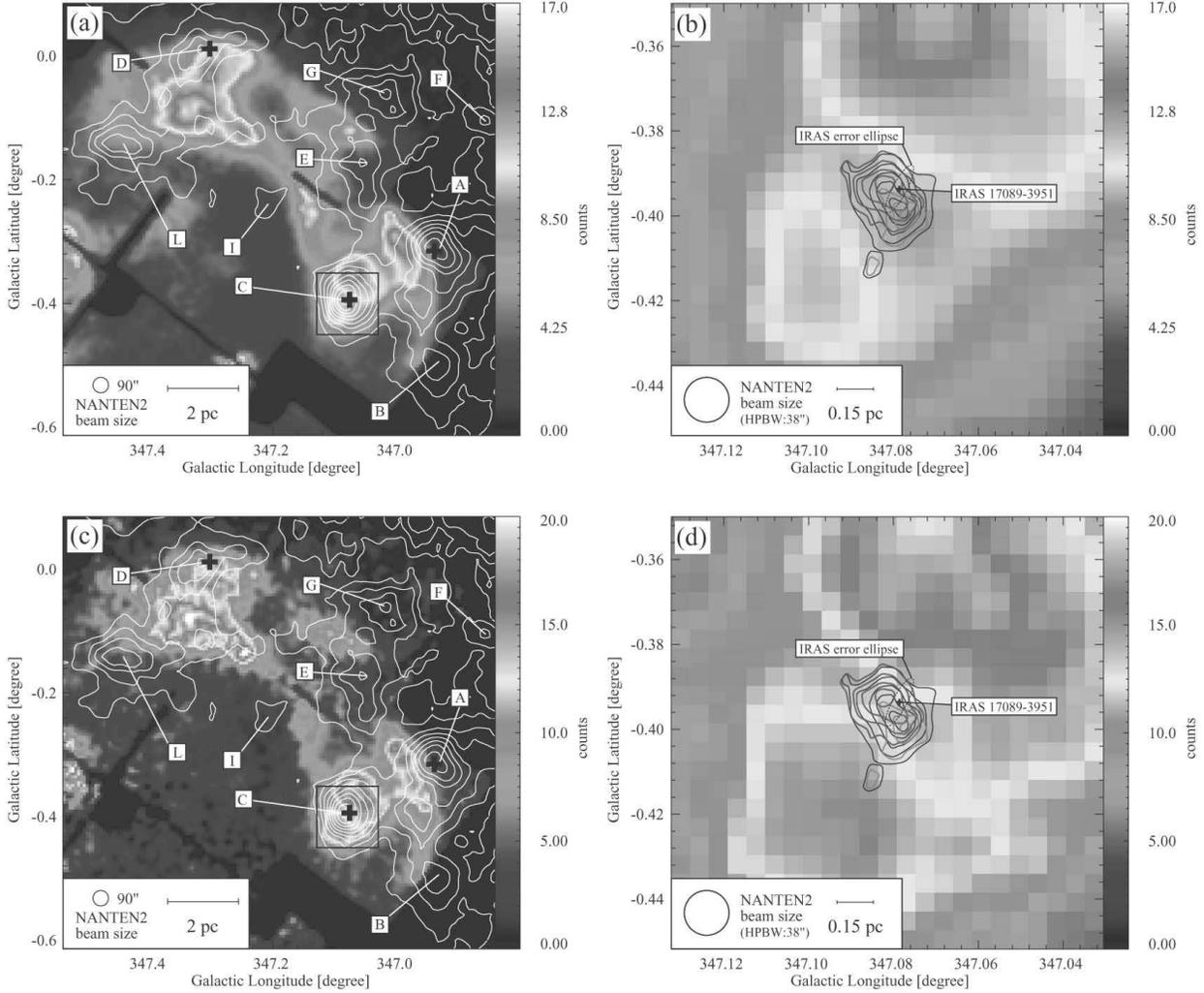}
\caption{$Suzaku$ XIS (XIS 0+2+3) mosaic image of RX J1713.7-3946 in the energy bands (a) 1-5 keV and (c) 5-10 keV in color scale (Tanaka et al. 2008, Figure 7) and $^{12}$CO($J$=2-1) intensity contours taken by NANTEN2. The contour levels and the velocity range of CO are the same as in Fig.1.(a). Open crosses show the positions of the $IRAS$ point sources (see Table 2.). The area enclosed by the black box is shown enlarged in Fig.5(b) and (d). (b) and (d) are $Suzaku$ XIS (XIS 0+2+3) mosaic image of peak C. The energy bands (b) 1-5 keV and (d) 5-10 keV in color scale are overlayed by contours of the $^{12}$CO($J$=4-3) core.}
\label{fig5}
\end{center}
\end{figure}%

\newpage
\begin{figure}[hbpt]
\begin{center}
\includegraphics[width=162mm,clip]{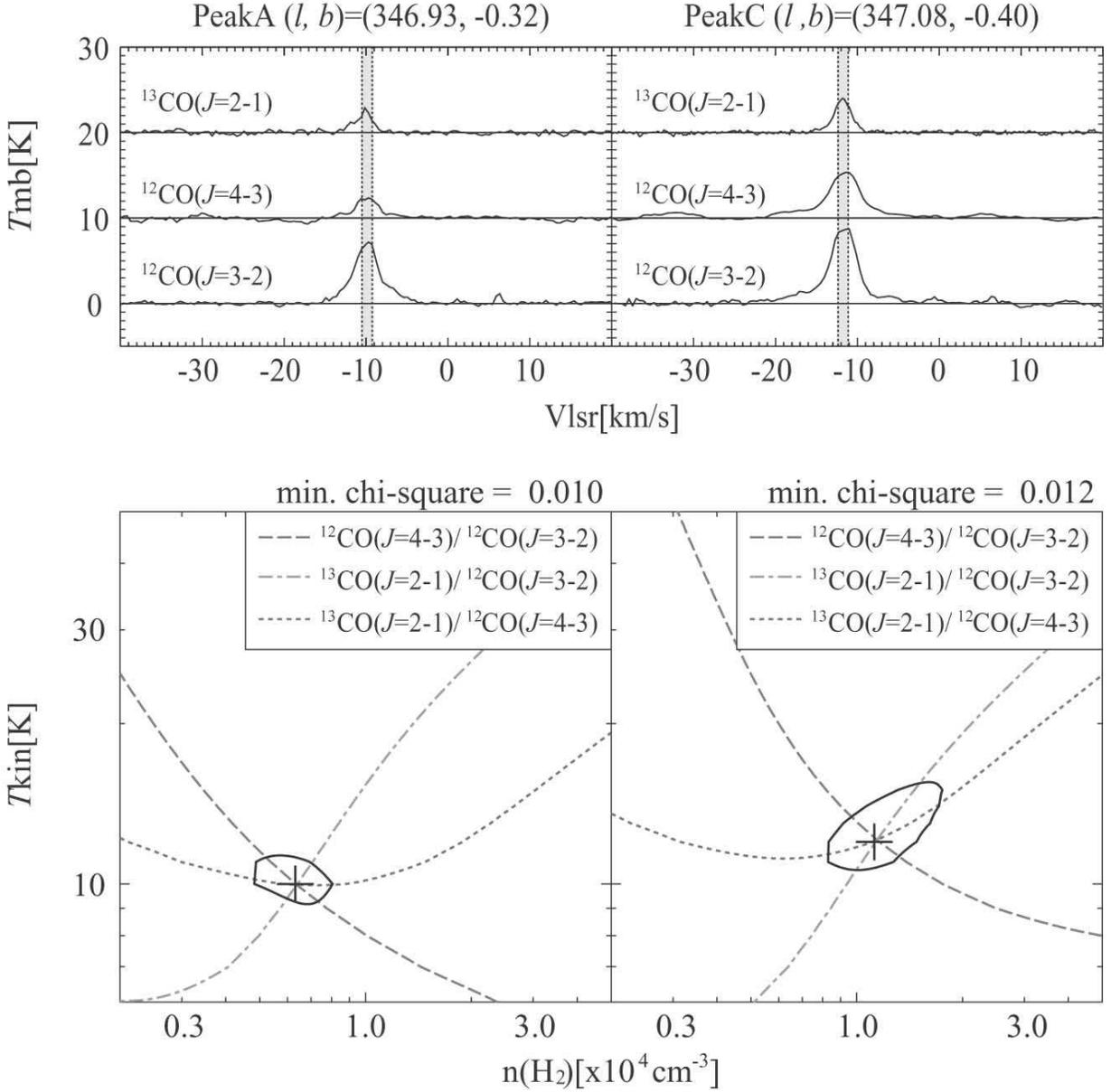}
\caption{Results of an LVG analysis for molecular peaks A and C. Top: CO line profiles for each molecular peak are smoothed to 90" by a Gaussian beam corresponding to the $^{13}$CO($J$=2-1) beam size. Bottom: Results of an LVG analysis for molecular peaks as indicated by the Top panels. Each locus of a black solid line surrounding the cross indicates the chi-square $\chi^2$ of 3.84, which corresponds to the 95$\%$ confidence level of a chi-square distribution with one degree of freedom. The crosses denote the lowest point of chi-square.
}
\label{fig6}
\end{center}
\end{figure}%

\newpage
\begin{figure}[hbpt]
\begin{center}
\includegraphics[width=100mm,clip]{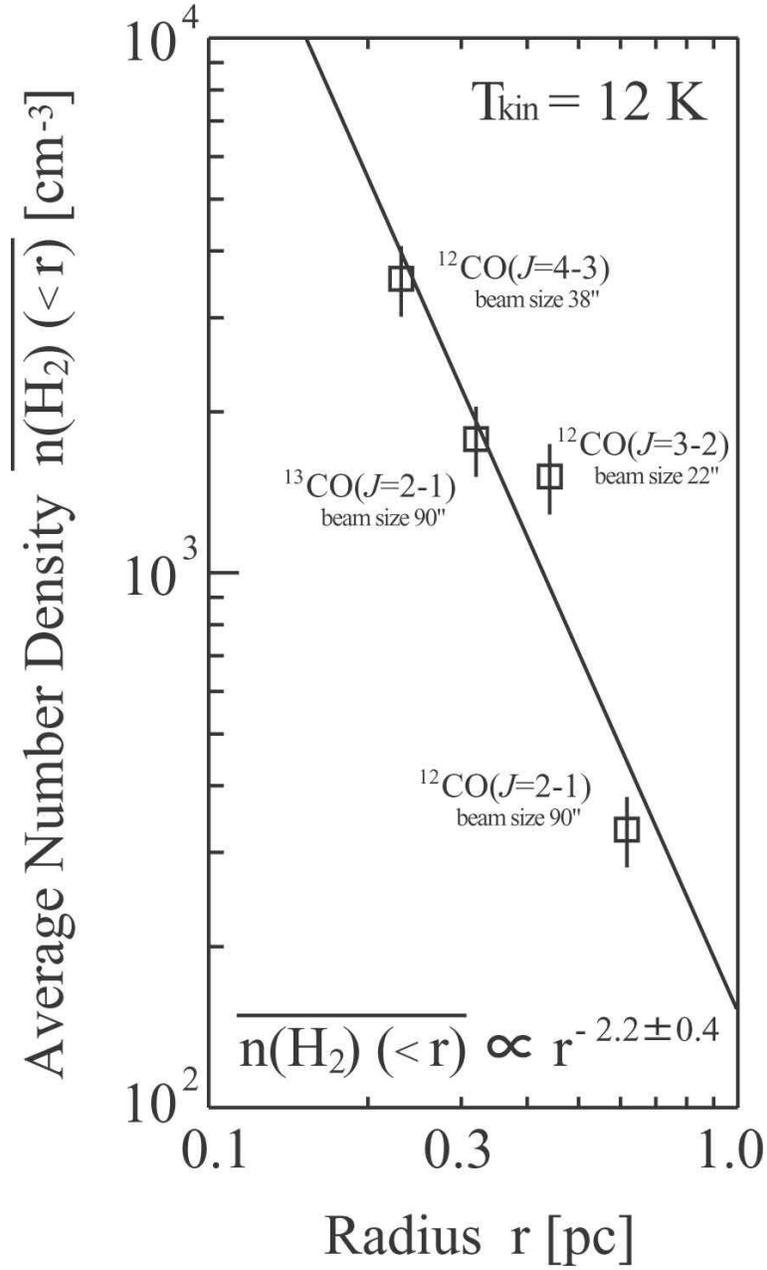}
\caption{The molecular density distribution of peak C. The calculations are carried out for $T_\mathrm{kin}$ = 12 K (see Table 3.). The solid line is a regression line obtained by least squares fitting.  Error bars represent only those for the radiation temperature calibration,$\pm 15\%$. See Table 4. for further.}
\label{fig7}
\end{center}
\end{figure}%

\begin{deluxetable}{cccccccccc}
\tablecaption{Properties of $^{12}$CO($J$=1-0) Clouds}
\tablewidth{0pt}
\tablehead{
Name & $l$ & $b$ & $T_R^\ast $ & $V_{\mathrm{peak}}$ & $\Delta V_{\mathrm{LSR}}$ & Size & $L_{\mathrm{CO}}$ & Mass &$IRAS$ \\
 & (deg) & (deg) & (K) & (km $\mathrm{s^{-1}}$) & (km $\mathrm{s^{-1}}$) & (pc) & \scalebox{0.81}[1]{($10^2$ K km $\mathrm{s^{-1}}$ $\mathrm{pc^2}$)} & ($M_\sun $) & point sources \\
(1) & (2) & (3) & (4) & (5) & (6) & (7) & (8) & (9) & (10) \\
}
\startdata
A & 346.933 & -3.000 & 8.5 & -10.3 & 4.8 & 4.2 & 32.5 & 686 
& 17082-3955  \\
B & 346.933 & -5.000 & 4.2 & \phantom{0}-8.0 & 4.6 & 2.1 & 9.0 & 190 & \nodata \\
C & 347.067 & -0.400 & 9.4 & -12.0 & 3.8 & 3.0 & 18.8 & 397 
& 17089-3951  \\
D & 347.300 & \phantom{0}0.000 & 4.0 & -10.1 & 4.8 & 3.0 & 13.9 & 292 
&  17079-3926   \\
E & 347.033 & -0.200 & 2.0 & \phantom{0}-6.1 & 7.2 & 3.0 & 7.5 & 159 &   \nodata \\
F & 346.867 & -0.100 & 2.3 & \phantom{0}-3.5 & 5.0 & 2.7 & \phantom{0}6.2 & 131 &  \nodata \\
G & 347.033 & -0.067 & 3.3 & -10.8 & 8.0 & 2.7 & 14.6 & 307 &  \nodata\\
I & 347.200 & -0.233 & 1.8 & \phantom{0}-9.9 & 5.4 & 2.4 & \phantom{0}4.9 & 103 &  \nodata \\
L & 347.433 & -0.133 & 4.0 & -12.0 & 5.7 & 3.0 & 17.6 & 370 & \nodata \\
\enddata
\label{tab1}
\tablecomments{Col. (1): Cloud name. Cols. (2)--(3): Position of the observed point with the maximum $^{12}$CO($J$=1-0) intensity. Cols. (4)--(6): Observed properties of the $^{12}$CO($J$=1-0) spectra obtained at the peak positions of the CO clouds. Col. (4): Peak radiation temperature $T_R^{\ast} $. Col. (5): $V_{\mathrm{LSR}}$ derived from a single Gaussian fitting. Col. (6): FWHM line width $\bigtriangleup V_{\mathrm{peak}}$. Col. (7): Size defined as ($A$/$\pi$)$^{0.5} \times 2$, where $A$ is the total cloud surface area defined as the region surrounded by the contour of 8.5 or 6.5 K (see text). If the contour is unclosed, the boundary is defined as the intensity minimum between the nearby peaks. Col. (8): The CO luminosity of the cloud $L_{\mathrm{CO}}$. Col. (9): Mass of the cloud derived by using the relation between the molecular hydrogen column density $N$($\mathrm{H_2}$) and the $^{12}$CO($J$=1-0) intensity $W$($^{12}$CO), $N$($\mathrm{H_2}$) = 2.0 $\times $  $10^{20}$[$W$($^{12}$CO)/(K km $\mathrm{s^{-1}}$)] ($\mathrm{cm^{-2}}$) ( Bertsch et al. 1993). Col. (10): $IRAS$ point source name nearby $^{12}$CO($J$=3-2) peaks.(see Cols. (1)--(9) are Moriguchi et al. 2005, Table 1)}
\end{deluxetable}

\clearpage



\begin{deluxetable}{cccccccccccccc}
\tabletypesize{\scriptsize}
\rotate
\tablecaption{Properties of IRAS point sources}
\tablewidth{0pt}
\tablehead{
Cloud & $IRAS$ & $l$ & $b$ & $\alpha$(J2000) & $\delta$(J2000) & semimaj & semimin & Posang &$F_{\mathrm{12}}$ & $F_{\mathrm{25}}$ & $F_{\mathrm{60}}$ & $F_{\mathrm{100}}$ & $L_{\mathrm{IRAS}}$ \\
name & point sources & (deg) & (deg) & ($\phantom{0}^{h}\phantom{0}^{m}\phantom{0}^{s}$) &  ($\phantom{0}\arcdeg \phantom{0}\arcmin \phantom{0}\arcsec$) & $\arcmin$ & $\arcmin$ & deg.& (Jy) & (Jy) & (Jy) & (Jy) & ($L_{\sun}$)  \\
(1) & (2) & (3) & (4) & (5) & (6) & (7) & (8) & (9) & (10) & (11) &(12) & (13) &(14) \\
}
\startdata
A & 17082-3955 & 346.94 & -0.31 & 17 11 41.04 & -39 59 11.21 & 23 & 5 & 98
& 5.4 & \phantom{0}3.8 & 17.5 & 138 & 137 \\
C & 17089-3951 &  347.08 & -0.39 & 17 12 26.46 & -39 55 17.98 & 23 & 6 & 98 
& 4.4 & 13.0 & 98.5 & 234 & 311 \\
D & 17079-3926 & 347.31 & \phantom{0}0.01 & 17 11 25.59 & -39 29 53.29 & 39 & 6 & 98 & 2.0 & 20.0 & 88.6 & 739 & 562 \\
\enddata
\label{tab2}
\tablecomments{Cols. (1)--(2): Cloud name (Moriguchi et al. 2005) and $IRAS$ point source near the $^{12}$CO($J$=3-2) peaks. Cols. (3)--(6): Position of the $IRAS$ sources. Cols. (7)--(9): Semimajour axis, semiminor axis, and position angle of the position error of the $IRAS$ sources. Cols (10)--(13): Fluxes of 12, 25, 60, and 100 $\mathrm{\mu m}$, respectively. Col. (14): $IRAS$ luminosity estimated using formula of Emerson (1988). Col. (18).(see Cols. (2)--(4) and (10)--(15) are Moriguchi et al. 2005, Table 3)}
\end{deluxetable}

\clearpage


\begin{deluxetable}{cccccccccc}
\tablecaption{Results of LVG Analysis at Molecular Peaks}
\tablewidth{0pt}
\tablehead{
 & & & \multicolumn{2}{c}{$\mathrm{^{12}CO}$} & &$\mathrm{^{13}CO}$ & & \\
\cline{4-5}\cline{7-7}
Name & $l$ & $b$ & $J$=3-2 & $J$=4-3 &&$J$=2-1 & $n$(H${_2}$) & $T_{\mathrm{kin}}$ \\
 & (\arcdeg) & (\arcdeg) & ($K$) & ($K$) &&($K$) & ($\mathrm{10^{4} cm^{-3}}$) & (K) \\
(1) & (2) & (3) & (4) & (5) && (6) & (7) & (8)  \\
}
\startdata
A & 346.94 & -0.32 & 6.6 & 2.1 && 1.9 & $0.6^{+0.2}_{-0.1}$ & $10^{+1}_{-1}$ \\
C & 347.08 & -0.40 & 8.5 & 5.2 && 3.4 & $1.1^{+0.6}_{-0.3}$ & $12^{+4}_{-1}$ \\
\enddata
\label{tab3}
\tablecomments{Col. (1): Cloud name. Col. (2)--(3): Position of the observed point with the maximum $\mathrm{^{12}CO}$($J$=3-2) intensity peak. Col. (4)-(6): Radiation temperature averaged to the line of sight over a velocity integral of 1.5 km s$^{-1}$. Col. (7): Density of molecular hydrogen. Col. (8): Kinetic temperature. The parameter, $X/$($dV/dR$)$ = 4.0 \times 10^{-6}$ (km s$^{-1}$ pc$^{-1}$)$^{-1}$, is used.}
\end{deluxetable}

\clearpage
\begin{deluxetable}{lcccccc}
\tablecaption{Density Distribution and Core Radius for Peak C}
\tablewidth{0pt}
\tablehead{ & & $\mathrm{^{12}CO}$ & & &$\mathrm{^{13}CO}$& \\
\cline{2-4}\cline{6-6}
Property & $J$=2-1 & $J$=3-2 & $J$=4-3 & &$J$=2-1 &}
\startdata
Beam size (arcsec) & 90 & 22 & 38 && 90 &\\
Core radius (pc) & 0.62 & 0.44 & 0.23 && 0.32& \\
Average brightness (K) & 0.79 & 1.87 & 1.28 && 0.29& \\
Number density ($\mathrm{10^3 cm^{-3}}$) & 0.33$\pm $0.05 & 1.51$\pm $0.23 & 3.55$\pm $0.53 && 1.78$\pm $0.27 & \\  
\enddata
\label{tab4}
\tablecomments{Number density is derived by assuming $T_{\mathrm{kin}} = 12$ K and $X/$($dV/dR$)$ = 4.0 \times 10^{-6}$ (km s$^{-1}$ pc$^{-1}$)$^{-1}$. We assume that the error of density caused by calibration from $T_{\mathrm{A}}^{\ast }$ to $T_{\mathrm{MB}}$ is $\pm $15$\%$ for all intensities. }
\end{deluxetable}

\clearpage

\begin{deluxetable}{lcc}
\tablecaption{Outflow Properties at Peak C in $^{12}$CO($J$=4-3)}
\tablewidth{0pt}
\label{table3}
\tablehead{Property & Blue & Red\\
 & component & component}
\startdata
Integrated intensity (K km $\mathrm{s^{-1}}$) & 17.5\phantom{0} & 14.7\phantom{0} \\
Size (arcmin) & 0.85 & 0.58 \\
Size (pc) & 0.25 & 0.17 \\
$\Delta V$ (km $\mathrm{s^{-1}}$) & 15 & 15\\
$t_{\mathrm{dyn}}$ ($10^4$ yr) & 1.6 & 1.1 \\
\enddata
\label{tab5}
\tablecomments{
Integrated intensity is derived by summing the integrated intensities of the observed points in the area enclosed by a contour of 13.4 K within the velocity range of $-30$ km s$^{-1} \leq V_{\rm LSR} \leq -15$ km s$^{-1}$ for blueshifted component and 11.0 K within $-8$ km s$^{-1} \leq V_{\rm LSR} \leq 7$ km s$^{-1}$ for redshifted component, respectively. Size is defined as an effective diameter=$\sqrt{(A/\pi)} \times 2$, where $A$ is the region enclosed by a contour of 13.4 K km s$^{-1}$ and 11.0 K km s$^{-1}$ for the blueshifted and redshifted component, respectively. $\Delta V$ is the velocity range of the wing component. The dynamical age, $t_{\mathrm{dyn}}$, is defined as  2$R/\Delta V$.}
\end{deluxetable}


\begin{thebibliography}{}
\bibitem[]{1} Aharonian, F. A., et al. 2004, Nature, 432, 75
\bibitem[]{2} Aharonian, F. A., et al. 2006, A\&A, 449, 223
\bibitem[]{3} Aharonian, F. A., et al. 2007, A\&A, 464, 235
\bibitem[]{6} Bertsch, D. L., Dame, T. M., Fichtel, C. E., Hunter, S. D., Sreekumar, P., Stacy, J. G., \& Thaddeus, P. 1993, \apj, 416, 587
\bibitem[]{0000} Blake, G. A., Sutton, E. C., Masson, C. R., \& Phillips, T. G. 1987, \apj, 315, 621
\bibitem[]{4} Cassam-Chenai, G., Decourchelle, A., Ballet, J., Sauvageot, J, L., Dubner, G., \& Giacani, E. 2004, A\&A, 427, 199
\bibitem[]{000} Denoyer, L. K. 1979, ApJL, 232, 165
\bibitem[]{0002} Desai, K. M., et al. 2010, \aj, 140, 584
\bibitem[]{7} Inoue, T., Yamazaki, R., \& Inutsuka, S. 2009, \apj, 695, 825
\bibitem[]{9} Fukui, Y. 1989, in Proc. ESO Workshop on Low Mass Star Formation and Pre-Main Sequence Objects, ed. B. Reipurth (ESO: Garching), 95
\bibitem[]{10} Fukui, Y., Iwata, T., Mizuno, A., Bally, J., \& Lane, A. P. 1993, in Protostars and Planets III, ed. E. H. Levy \& J. I. Lunine (Tucson: Univ. Arizona Press), 603
\bibitem[Fukui et al. (2003)]{fukui03} Fukui, Y., et al. 2003, \pasj, 55, 61
\bibitem[Fukui et al. (2008)]{fukui08} Fukui, Y. 2008, in AIP Conf. Proc., Vol. 1085, Proc. of 4th International Meeting on High-Energy Gamma-Ray Astronomy, ed. F. A. Aharonian, W. Hofmann, \& F. Rieger (Melville, NY: AIP), 104
\bibitem[]{12} G$\mathrm{\ddot{u}}$sten, R., \& Philipp, S. D. 2004, in Proc. the Fourth Cologne-Bonn-Zermatt Symposium ed. S. Pfalzner, C. Kramer, C. Staubmeier, \& A. Heithausen (Heidelberg: Springer), 253
\bibitem[]{11} Goldreich, P., \& Kwan, J. 1974, \apj, 189, 441
\bibitem[]{0001} Koo, B.-C., et al. 2008, ApJL, 673, 147
\bibitem[Koyama et al. (1997)]{Koyama97} Koyama, K., Kinugasa, K., Matsuzaki, K., Nishiuchi, M., Sugizaki, M., Torii, K., Yamauchi, S., \& Aschenbach, B. 1997, PASJ, 49, 7
\bibitem[]{14} Kulesa, C. A., Hungerford, A. L., Walker, C. K., Zhang, X., \& Lane, A. P. 2005, \apj, 625, 194
\bibitem[]{16} Lada, C. J. 1986, S\&T, 72, 334
\bibitem[]{17} Larson, R. B. 1969, MNRAS, 145, 271
\bibitem[]{18} Lizano, S., \& Shu, F. H. 1989, \apj, 342, 834
\bibitem[Mizuno and Fukui (2004)]{mizuno04} Mizuno, A., \& Fukui, Y. 2004, in ASP Conf. Ser. 317, Milky Way Surveys: The Structure and Evolution of our Galaxy, ed. D. Clemens, R. Shah, \& T. Brainerd (San Francisco, CA: ASP), 59
\bibitem[]{mizuno10} Mizuno, Y., et al. 2010, PASJ, 62, 51
\bibitem[]{00} Moriguchi, Y., Tamura, K., Tawara, Y., Sasago, H., Yamaoka, K., Onishi, T., \& Fukui, Y. 2005, \apj, 631, 947
\bibitem[]{21} Onishi, T., et al. 1999, PASJ, 51, 871
\bibitem[]{22} Penston, M. V. 1969, MNRAS, 145, 457
\bibitem[Pfeffermann \& Aschenbach (1996)]{Pfeffermann96} Pfeffermann, E., \& Aschenbach, B. 1996, in Proc. R$\mathrm{\ddot{o}}$ntgenstrahlung from the Universe, ed. H. U. Zimmermann, J. Tr$\mathrm{\ddot{u}}$mper, \& H. Yorke (MPE Rep. 263; Garching: MPE), 267
\bibitem[]{24} Pineda, J. L., et al. 2008, A\&A, 482, 197
\bibitem[]{26} Scoville, N. Z., \& Solomon, P. M. 1974, ApJL, 187, 67
\bibitem[Slane et al. (1999)]{Slane99} Slane, P., Gaensler, B. M., Dame, T. M., Hughes, J. P., Plucinsky, P. P., \& Green, A. 1999, \apj, 525, 357
\bibitem[]{5} Tanaka, T., et al. 2008, \apj, 685, 988
\bibitem[]{0003} Truelove, J. K., \& McKee, C. F. 1999, ApJS, 120, 299
\bibitem[]{29} Uchiyama, Y., Aharonian, F. A., Tanaka, T., Takahashi, T., \& Maeda, Y. 2007, Nature, 449, 576
\bibitem[]{30} Wang, Z. R., Qu, Q.-Y., \& Chen, Y. 1997, A\&A, 318, 59
\bibitem[]{31} Wootten, H. A. 1977, \apj, 216, 440
\bibitem[]{32} Wootten, A. 1981, \apj, 245, 105
\end{thebibliography}
\end{document}